\documentclass[aps,pra,superscriptaddress,notitlepage,10pt,a4paper,twocolumn,floatfix]{revtex4-1}

\usepackage[pdftex]{graphicx}
\usepackage{amsmath}
\usepackage{hyperref}
\usepackage{times}
\usepackage[centering,hmargin=2cm,vmargin=2.5cm]{geometry}

\begin{document}
\title{Efficient saturation of an ion in free space}

\author{Martin Fischer}
\email{martin.fischer@mpl.mpg.de}
\affiliation{Institute of Optics, Information and Photonics, University of Erlangen-Nuremberg, 91058 Erlangen, Germany}
\affiliation{Max Planck Institute for the Science of Light, Guenther-Scharowsky-Str. 1/Bldg. 24, 91058 Erlangen, Germany}
\affiliation{These authors contributed equally to this work.}

\author{Marianne Bader}
\affiliation{Institute of Optics, Information and Photonics, University of Erlangen-Nuremberg, 91058 Erlangen, Germany}
\affiliation{Max Planck Institute for the Science of Light, Guenther-Scharowsky-Str. 1/Bldg. 24, 91058 Erlangen, Germany}
\affiliation{These authors contributed equally to this work.}

\author{Robert Maiwald}
\affiliation{Institute of Optics, Information and Photonics, University of Erlangen-Nuremberg, 91058 Erlangen, Germany}
\affiliation{present address: Physikalisches Institut, University of Bonn, 53115 Bonn, Germany}

\author{Andrea Golla}
\affiliation{Institute of Optics, Information and Photonics, University of Erlangen-Nuremberg, 91058 Erlangen, Germany}
\affiliation{Max Planck Institute for the Science of Light, Guenther-Scharowsky-Str. 1/Bldg. 24, 91058 Erlangen, Germany}

\author{Markus Sondermann}
\email{markus.sondermann@fau.de}
\affiliation{Institute of Optics, Information and Photonics, University of Erlangen-Nuremberg, 91058 Erlangen, Germany}
\affiliation{Max Planck Institute for the Science of Light, Guenther-Scharowsky-Str. 1/Bldg. 24, 91058 Erlangen, Germany}

\author{Gerd Leuchs}
\affiliation{Institute of Optics, Information and Photonics, University of Erlangen-Nuremberg, 91058 Erlangen, Germany}
\affiliation{Max Planck Institute for the Science of Light, Guenther-Scharowsky-Str. 1/Bldg. 24, 91058 Erlangen, Germany}


\begin{abstract}
\label{abstract}

We report on the demonstration of a light-matter interface coupling light to a single $^{174}\textrm{Yb}^+$ ion in free space. The interface is realized through a parabolic mirror partially surrounding the ion. It transforms a Laguerre-Gaussian beam into a linear dipole wave converging at the mirror's focus. By measuring the non-linear response of the atomic transition we deduce the power required for reaching an upper-level population of $1/4$ to be $692\pm20~\textrm{pW}$ at half linewidth detuning from the atomic resonance. Performing this measurement while scanning the ion through the focus provides a map of the focal intensity distribution. From the measured power we infer a coupling efficiency of $7.2\pm0.2~\%$ on the linear dipole transition when illuminating from half solid angle, being among the best coupling efficiencies reported for a single atom in free space.

\end{abstract}

\maketitle

\section{Introduction}
\label{intro}

Coupling light and matter is an essential part in many quantum information protocols \cite{RevModPhysGisin2002}. For many of these protocols to be implemented successfully this coupling should be as high as possible. The scheme used most frequently for achieving high coupling efficiencies is to place the matter system into a high-quality resonator. Here we will rather focus on light-matter coupling in free space and the measurement and characterization of the coupling efficiency. When investigating this efficiency typically three distinct effects are measured:

Firstly, efficient coupling increases the probability of a photon being absorbed by a matter system. This provides an opportunity to use matter as a quantum memory in which one can store the state of a photon. Here, a measure for the coupling efficiency could be the probability with which one can store a single photon in a matter system \cite{NatPhysPiro2011}. As for all other types of experiments described below, spatially mode matching the light field to the transition is essential. In addition, depending on the inner structure of the matter system involved, it is necessary to create an optimal temporal shape of the incident field \cite{OptCommQuabis2000,PhyScrLeuchs2012,PhysRevLettAljunid2013}. Thus the absorption probability is affected by two different effects that have to be distinguished by additional measurements. 

A second effect occurring in light-matter interaction is the phase shift that a light field acquires when interacting dispersively with a medium. Here, the phase a field accumulates in comparison to a non-interacting field provides a good measure for the strength of the interaction \cite{JEurOptSocRapPublicSondermann2013,PhysRevLettPotoschnig2011,PRLAljunid2009,PRAHetet2013}, given the light is scattered coherently. This, however, is only the case if no upper-level population is induced and hence there is no incoherent scattering. This is only the case if the driving field is zero. To account for the amount of incoherently scattered light the upper-level population has to be determined. Additionally, investigating the phase of a field requires some way of stabilizing the phase of the non-interacting field.

A further effect that is often investigated in the context of high coupling efficiencies is the extinction of a light field traveling past a matter system. This effect is related to the previous one, because both effects originate from the interference between the impinging light and the light scattered by the emitter \cite{PRLZumofen2008,PRLAljunid2009,PhysRevLettPotoschnig2011}. In contrast to the phase shift, which is maximized by illumination from full solid angle \cite{JEurOptSocRapPublicSondermann2013,JourModOptLeuchs2012}, optimal results are obtained when focusing light from half the solid angle \cite{PRLZumofen2008}. The depth of the dip in the transmission through the system provides a good measure for the amount of light that was interacting \cite{PhysRevAKochan1994,NatPhysWrigge2008,NJPTey2009,PRLSlodicka2010}. Like in the case of the phase shift the effect is reduced by incoherently scattered light \cite{NatPhysWrigge2008}.

Here, we will establish saturation measurements as a tool for characterizing the coupling efficiency in free space in an unambiguous way, utilizing the very effect that is detrimental in the types of measurements discussed above. The next section discusses the advantages of saturation measurements in more detail and reviews the relation between coupling efficiency and the necessary power to reach a given upper-level population. The experimental set-up is described in Sec. \ref{setup}, whereas the experimental results are presented in Sec. \ref{results} and discussed in Sec. \ref{discussion}.

\section{Saturation Measurements}
\label{saturation}

In what follows we present an approach that provides a measure for the spatial overlap of the light field with the driven transition while neglecting temporal effects. A two-level-system \mbox{(TLS)} responds to the power of the driving field in a non-linear way. The amount of light scattered by a TLS is directly proportional to its upper-level population $\rho$. Solving the Bloch equations one finds that for strong driving fields the upper-level population in the steady state solution asymptotically reaches $\rho=1/2$ where the TLS scatters at a rate of $\Gamma/2$, where $\Gamma$ is the spontaneous emission rate of the TLS. Thus, one can directly relate the upper-level population to the amount of scattered photons. For example, at an upper-level population of $\rho=1/4$ the TLS scatters at a rate of $\Gamma/4$. This value is commonly associated with a saturation parameter $S=1$ in the literature by $\rho=\frac{S}{2(S+1)}$. Following Ref. \cite{PRAvanEnk2004} the rate at which the ion scatters photons equals

\begin{eqnarray}
\textrm{R}_\textrm{sc} = \frac{\Gamma}{2}\left(1-\frac{1+\delta^2}{1+\delta^2+8\left|\beta\right|^2/\Gamma}\right)
\label{eq:scat}
\end{eqnarray}  

with $|\beta|^2$ the rate of dipolar photons, $\delta=2\Delta/\Gamma$ with $\Delta$ the detuning from resonance. However, a real beam is in general comprised of dipolar and non-dipolar parts. Hence the rate of interacting photons is lowered by the ratio of dipolar parts. This ratio can be expressed by the coupling efficiency  which characterizes the focusing geometry including the overlap of the incident radiation pattern with the dipole transition of the TLS. In Refs. \cite{JourModOptLeuchs2012,EurPhysJGolla2012} the coupling efficiency has been defined as $G=\Omega\eta^2$, where $\Omega$ denotes the solid angle fraction of the focusing geometry weighted with the dipolar emission pattern. The overlap with the dipole is accounted for by $\eta$ which may also account for distortions of the phase front of the incident beam. In addition, there might be some other potentially unknown sources of imperfection influencing the coupling efficiency like e.g. residual motion \cite{OptCommTeo2011}. We will include these by a loss factor $1-L$ resulting in $G=\Omega\eta^2\left(1-L\right)$. Thus, a scattering rate of $\textrm{R}_\textrm{sc}=\Gamma/4$ and an upper-level population of $\rho=1/4$ is reached at an impinging photon rate of

\begin{eqnarray}
|\tilde{\beta}|^2=\frac{\Gamma}{8}\left(1+\delta^2\right)/G
\label{eq:ratein}
\end{eqnarray}

In the optimal case all incident photons interact with the TLS and $G=1$. From this the minimal power to reach an upper-level population of $\rho=1/4$ can be calculated to be $P_{\rho=1/4}=\hbar \omega_0 \Gamma /8$ on resonance \cite{PhysRevAKochan1994}, with the atomic transition frequency $\omega_0$.

By varying the incident power and evaluating \mbox{Eq. (\ref{eq:scat})} one can deduce the necessary power $P^{\textrm{exp}}_{\rho=1/4}$ at which this upper-level population is reached. Comparing the minimal power and incident power measured in the experiment results in the coupling efficiency

\begin{eqnarray}
G=P_{\rho=1/4}/P^{\textrm{exp}}_{\rho=1/4}
\label{eq:G_spa}
\end{eqnarray}

This way of measuring the coupling efficiency has several advantages: First of all it is independent of the losses in the detection system since they can be factored out by normalizing the rate of detected photons when strongly driven to its asymptotic value at full saturation. Furthermore, compared to measurements that are restricted to the weak excitation regime which is often noise-dominated, this is a fast method if the overall detection efficiency is sufficiently high. Finally it is not subject to detrimental effects based on the saturation of the upper-level population since these are the very effects that are utilized in this method.

\section{Experimental Set-up}
\label{setup}
\begin{figure} [Htb]
	\centering
	\resizebox{0.95\columnwidth}{!}{
		\includegraphics{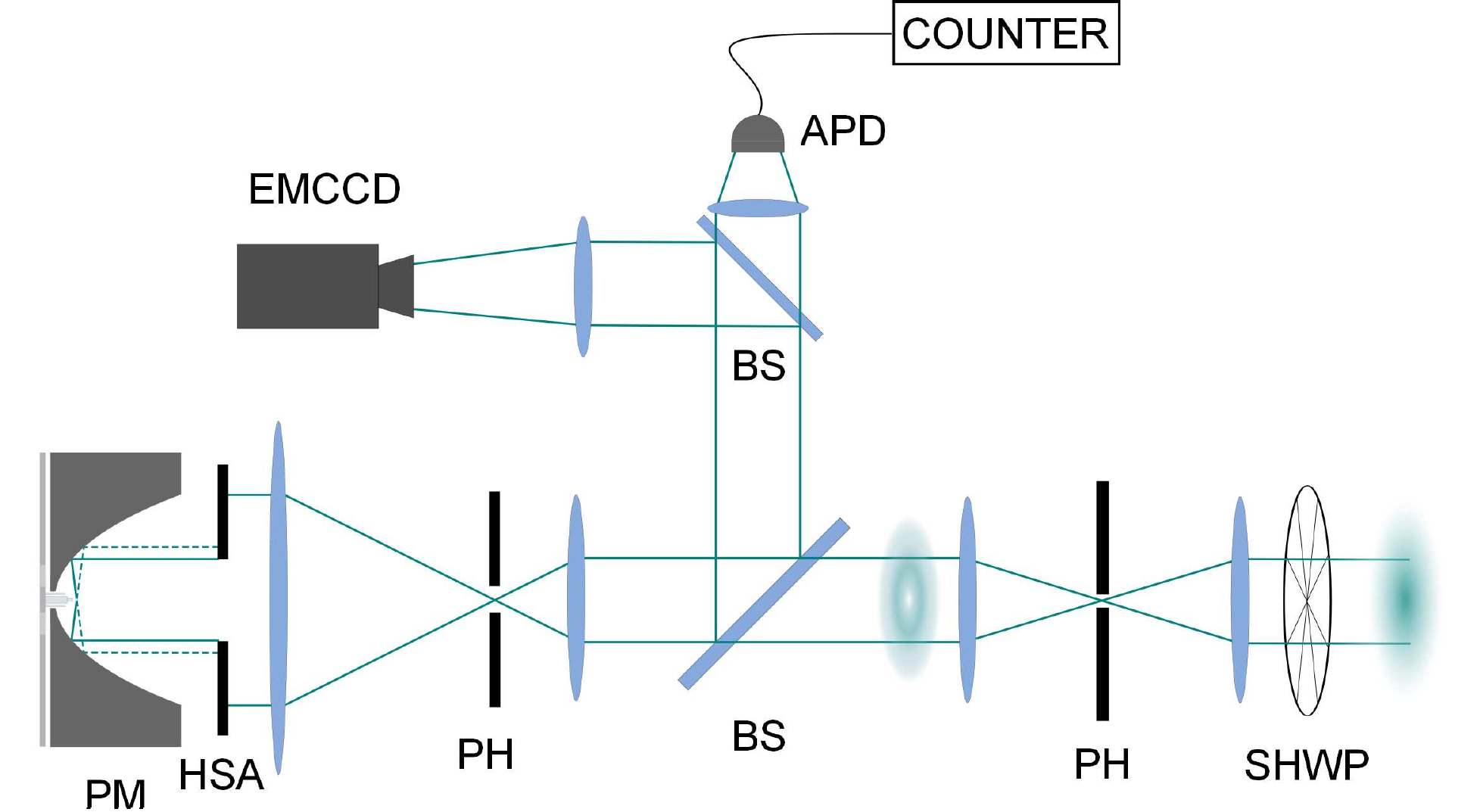}
		}
	\caption{Schematic of the set-up. The Laguerre-Gaussian beam is created by sending a Gaussian beam onto a segmented half wave plate \mbox{(SHWP)} and subsequent filtering by a pinhole \mbox{(PH)}. It is then transmitted to the parabolic mirror \mbox{(PM)} through a 50/50-beamsplitter \mbox{(BS)}. In order to be able to distinguish the light scattered by the ion from the driving laser field the beam is cut by an aperture \mbox{(HSA)} cutting the beam to half of the solid angle. Solid lines indicate the extreme most parts of the beam going towards the ion. The dashed lines show the path of the light after interaction with the ion. The light scattered by the ion is simultaneously monitored by an EMCCD and an APD.}
	\label{fig:setup}
\end{figure}

Fig. \ref{fig:setup} shows a schematic of the experimental set-up. The basic arrangement has been described earlier in Ref. \cite{PhysRevAMaiwald2012}. At  the heart of the experiment lies an ion trap similar to that used in Ref. \cite{NatPhysMaiwald2009} and a parabolic mirror of focal length $\textit{f} = 2.1~\textrm{mm}$. For the saturation measurements performed here this mirror serves two purposes: It collects the light scattered by the ion and thus is the first element of the detection system. Additionally it serves as a mode converter that transforms an incoming first order Laguerre-Gaussian mode with radial polarization into a linear dipole wave converging at the focus\cite{LaserPhysLindlein2007,ApplPhysBSondermann2007}. This light mode is generated by sending a linearly polarized Gaussian beam onto a segmented half wave plate \mbox{(SHWP)}. The resulting beam is spatially filtered by a 30-$\mu \textrm{m}$ pinhole such that only the two lowest order modes are transmitted \cite{EurPhysJGolla2012}. The size of the Laguerre-Gaussian beam is adjusted such that it yields the maximal overlap with a linear dipole field. Afterwards part of the beam is transmitted through a non-polarizing beamsplitter to the mirror.

In order to control the position of the ion relative to the focus of the parabolic mirror the trap is mounted on a linear \textit{x-y-z}-piezo stage. The correct position is found by imaging the light scattered by the ion onto an electron-multiplying charge-coupled device \mbox{(EMCCD)}. Afterwards the ion is fine positioned by maximizing the amount of light scattered by the ion when driven by the Laguerre-Gaussian mode.
In order to distinguish the light driving the ion from the scattered light it is necessary to filter the impinging laser light out of the detection path. In our set-up this is done by an aperture \mbox{(HSA)} with a diameter of $D=2\times\textit{f}$. Hence we illuminate the ion only from the inner part of the mirror which corresponds to half of the solid angle. After a first reflection on the parabolic mirror the inner half of the beam is focused onto the ion and then collimated again after a second reflection in the outer half of the mirror. In this way the excitation beam is blocked by the same aperture used to cut it in the first place. Due to the high collection efficiency provided by the mirror the ion can be simultaneously monitored by the \mbox{EMCCD} and by an avalanche photodiode \mbox{(APD)}.

To minimize losses originating from residual motion the correlation between the ion's fluorescence and the radio frequency applied to the trap is monitored \cite{JApplPhysBerkeland1998}. Since the Laguerre-Gaussian beam excites the ion from almost all directions it is possible, by using different apertures, to measure the correlation signal from three linearly independent directions. The excess micromotion is compensated by minimizing the modulation of the correlation signal for these directions.

In order to measure the couping efficiency $G$ the saturation of the $^{2}\textrm{S}_{1/2}\leftrightarrow$$^{2}\textrm{P}_{1/2}$ cooling transition at a wavelength of $369.5~\textrm{nm}$ of $^{174}\textrm{Yb}^+$ with a natural linewidth of $\Gamma/2\pi=19.6~\textrm{MHz}$ is investigated. Since the ion decays to the $^2D_{3/2}$ level with a probability of $0.5~\%$ \cite{PRAOlmschenk2007} we repump it with light at a wavelength of $935~\textrm{nm}$ to the $^3\textrm{D}[3/2]_{1/2}$ level from where it decays back into the cooling cycle \cite{PRABell1991}. The power of the Laguerre-Gaussian beam is varied by an acousto-optic modulator \mbox{(AOM)}. The scattered light is detected for $100~\textrm{ms}$. During this time the repumping laser is applied to avoid optical pumping to the meta-stable $^{2}\textrm{D}_{3/2}$ state. The $^{2}\textrm{D}_{3/2}\leftrightarrow$$^{3}\textrm{D}[3/2]_{1/2}$ transition is strongly driven to ensure that the ion can be treated as a close approximation of a pure TLS. Afterwards the repumping beam is switched off and the $369.5~\textrm{nm}$ transition is strongly driven for $10~\mu\textrm{s}$ to pump the ion into the $^{2}\textrm{D}_{3/2}$ state. Then the background light is measured for $100~\textrm{ms}$ and subtracted. In order to provide cooling during the experimental sequence the Laguerre-Gaussian beam is detuned by half a linewidth from resonance.

\section{Results}
\label{results}

\begin{figure} [Htb]
	\centering
	\resizebox{0.95\columnwidth}{!}{
		\includegraphics{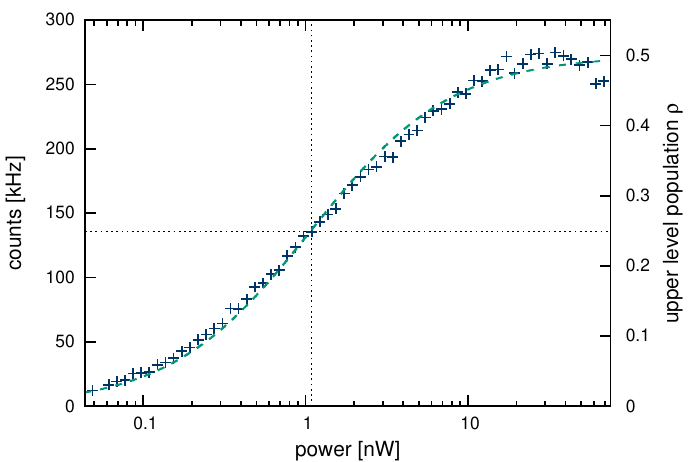}
		}
	\caption{Saturation curve of a single $^{174}\textrm{Yb}^+$ ion illuminated by a Laguerre-Gaussian beam from half solid angle. The data points are background corrected and were taken at half linewidth detuning from resonance. The dashed line shows a fit to the data, yielding a necessary power of $1081~\textrm{pW}$ incident onto the parabolic mirror to reach an upper-level population of $\rho=1/4$. Each data point was measured for $100~\textrm{ms}$.}
	\label{fig:saturation}
\end{figure}

Figure \ref{fig:saturation} shows the results of a typical measurement under the condition of optimal alignment. Fitting Eq. (\ref{eq:scat}) to the data yields a required power of $P^{\textrm{exp}}_{\rho=1/4}=1081~\textrm{pW}$ impinging on the mirror to reach an upper-level population of $\rho=1/4$. Factoring out the reflectivity of the mirror of $64~\%$ for a radially polarized Laguerre-Gaussian beam leads to a power of $692\pm20~\textrm{pW}$ impinging on the ion. 

For the $^{2}\textrm{S}_{1/2}\leftrightarrow$$^{2}\textrm{P}_{1/2}$ transition of $^{174}\textrm{Yb}^+$ Eq. (\ref{eq:ratein}) yields a minimal necessary power of $P_{\rho=1/4}=16.6~\textrm{pW}$ at a detuning of $\Delta=\Gamma/2$ to reach an upper-level population of $\rho=1/4$. However since in our experiment we are only aiming at driving the $\pi$-transition and not the $\sigma^\pm$-transitions this power has to be increased by a factor of 3 to $49.7~\textrm{pW}$ according to the Clebsch-Gordan coefficients of a $\textrm{J}=1/2$ to $\textrm{J}=1/2$ transition. Thus our system has a coupling efficiency of $G=7.2\pm0.2\%$. One has to keep in mind, that we are illuminating the ion only from one half of the solid angle, thus the maximally achievable coupling strength is $50~\%$ in this configuration. However, one has to account for known deficiencies. Interferometric measurements on the parabolic mirror \cite{ApplOptLeuchs2008} predict a Strehl ratio of $87~\%$. In combination with the measured overlap of the field incident on the parabolic mirror \cite{EurPhysJGolla2012} this yields an overlap of $\eta=0.91$. Accounting for the hole in the vertex of the parabolic mirror as well as the constraint to half solid angle leads to $\Omega=0.49$. This suggests an expected coupling efficiency $G=40.5~\%$. In order to investigate the origin of the discrepancy from the measured value, a scan of the focal intensity distribution is performed. 

The focus of the Laguerre-Gaussian beam inside the parabolic mirror is characterized utilizing the fact that the trap is mounted on a \textit{x-y-z}-piezo stage. Thus it is possible to move the ion through the focus and measure the necessary power to reach an upper-level population of $\rho=1/4$ at every point. The inverse of this power provides a quantity proportional to the local intensity of the electric field. Since moving the ion to different locations relative to the mirror's focus changes the detection efficiency of the set-up, the saturation measurements are a reliable tool as their outcome is independent of this efficiency.

\begin{figure} [Htb]
	\centering
	\resizebox{1.0\columnwidth}{!}{
		\includegraphics{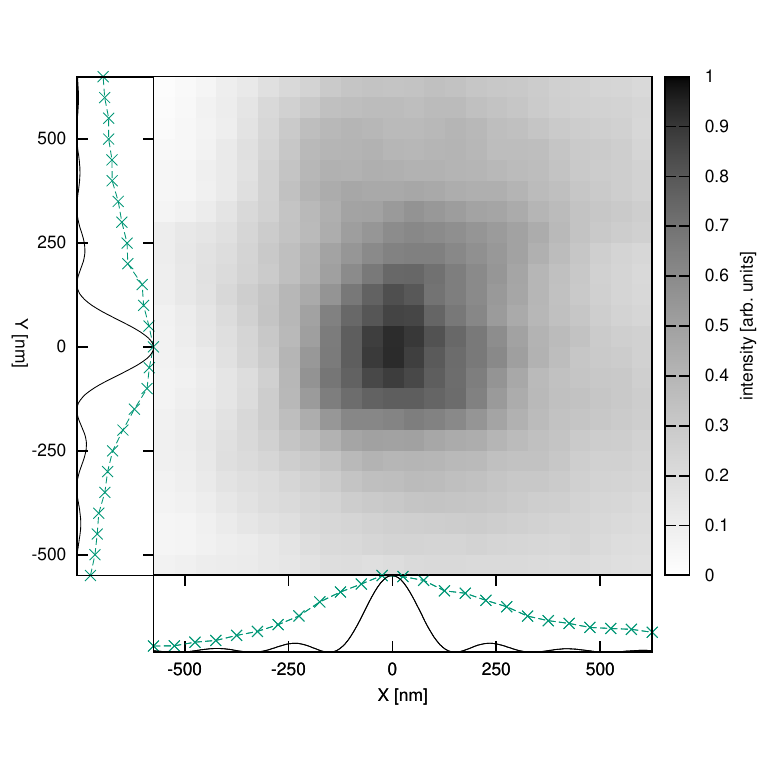}
		}
	\caption{Intensity distribution of the focus of the Laguerre-Gaussian beam after reflection off the parabolic mirror. The scan is taken perpendicular to the optical axis. The resolution is $50~\textrm{nm}/\textrm{pixel}$. Each pixel value is obtained by a single saturation measurement. The plots on the left and bottom side show a cut through the maximum of the focus. The solid black lines show a simulation of the intensity distribution accounting for the known aberrations of the parabolic mirror. The FWHM of the measured intensity distribution amounts to $530~\textrm{nm}$ in X-direction and $610~\textrm{nm}$ in Y-direction.}
	\label{fig:focus}
\end{figure}

Figure \ref{fig:focus} shows a scan through the focus perpendicular to the optical axis of the parabolic mirror. The scan is taken at a resolution of $50~\textrm{nm}/\textrm{pixel}$ and takes approximately seven minutes. This resolution is chosen to match the extent of the wave function of the ion assuming Doppler-limited cooling and considering the measured trap frequencies of $560~\textrm{kHz}$ in the radial directions. Analyzing the data yields a full width at half maximum \mbox{(FWHM)} of $530~\textrm{nm}$ in X-direction and $610~\textrm{nm}$ in Y-direction. Scanning the focus along the optical axis yields a FWHM of $660~\textrm{nm}$ in Z-direction.

From interferometric measurements performed on the parabolic mirror we calculate the expected shape of the intensity distribution in the focus when illuminated by a Laguerre-Gaussian beam from half solid angle. These simulations predict a FWHM of $140~\textrm{nm}$ perpendicular to the optical axis and $415~\textrm{nm}$ along the optical axis. Additional measurements of the phasefront and polarization of the Laguerre-Gaussian beam were performed. These measurements do not indicate that the focus should broaden significantly. Since the measured focal distribution is clearly larger than that expected from the simulations there seems to be a blurring effect. This blurring might originate in turning grooves that stem from the manufacturing process of the mirror.

\section{Discussion}
\label{discussion}

By spatially tailoring the incident light mode to the emission pattern of the linear dipole transition we are able to reach coupling strengths of approximately 7~\% while exciting the ion from half solid angle. Scanning the ion through the focus of the parabolic mirror shows that the extent of the focus is greater than what is to be expected from simulations by a factor of 4.1 in the transversal directions and by a factor of 1.6 in longitudinal direction. Since these factors are on the same order of magnitude as the deviation of the measured and the expected coupling efficiency, we conjecture that the blurring of the focus is responsible for this mismatch. Nevertheless, this coupling efficiency is, to our knowledge, among the best reported for single atoms in free space.

In other experiments, the coupling efficiency as defined here is seldom specified. An exception is Ref. \cite{PhysRevLettAljunid2013} in which a coupling efficiency of $3~\%$ is reported. For most of the other works on light-matter coupling in free space one has to estimate the coupling efficiency from the data provided. For Ref. \cite{NatPhysWrigge2008} based on an extinction of $22~\%$ we infer a coupling efficiency with a magnitude comparable to the one reported here.

One should, however, recall that our results were obtained without correcting for the aberrations of the parabolic mirror and a non-optimal surface quality. This suggests that the use of a mirror with better surface quality as well as aberration correction will boost the coupling efficiency to values on the order of $90~\%$ when illuminating the ion from full solid angle, as it has been envisioned in Ref. \cite{EurPhysJGolla2012} based on the quality of the incident optical mode. This would provide a powerful tool for many quantum information protocols as well as promising steps towards new quantum mechanical technologies such as e.g. a quantum transistor \cite{NatureHwang2009}.

\section{Acknowledgments}
\label{acknowledgments}
The authors thank S. Heugel for fruitful discussions. M. S. acknowledges financial support from the Deutsche Forschungsgemeinschaft (DFG). G. L. wants to thank the German Federal Ministry of Education and Research (BMBF) for financial support in the framework of the joint research project \textit{QuORep}.


\begin{thebibliography}{}
%
%
\bibitem{RevModPhysGisin2002}
N. Gisin, G. Ribordy, W. Tittel, H. Zbinden, Rev. Mod. Phys. \textbf{74}, 145-195 (2002) 
\bibitem{NatPhysPiro2011}
N. Piro, F. Rohde, C. Schuck, M. Almendros, J. Huwer, J. Ghosh, A. Haase, M. Hennrich, F. Dubin, J. Eschner, Nature Physics \textbf{7}, 17-20 (2011)
\bibitem{OptCommQuabis2000}
S. Quabis, R. Dorn, M. Eberler, O. Gl\"{o}ckl, G. Leuchs, Opt. Comm. \textbf{179}, 1-7 (2000)
\bibitem{PhyScrLeuchs2012}
G. Leuchs, M. Sondermann, Physica Scripta \textbf{85}, 058101 (2012)
\bibitem{PhysRevLettAljunid2013}
S. A. Aljunid, G. Maslennikov, Y. Wang, H. L. Dao, V. Scarani, C. Kurtsiefer, Phys. Rev. Lett. \textbf{111}, 103001 (2013)
\bibitem{JEurOptSocRapPublicSondermann2013}
M. Sondermann, G. Leuchs, J. Europ. Opt. Soc. Rap. Public. \textbf{8}, 13502 (2013)
\bibitem{PhysRevLettPotoschnig2011}
M. Pototschnig, Y. Chassagneux, J. Hwang, G. Zumofen, A. Renn, V. Sandoghdar, Phys. Rev. Lett. \textbf{107}, 063001 (2011)
\bibitem{PRLAljunid2009}
S. A. Aljunid, M. K. Tey, B. Chng, T. Liew, G. Maslennikov, V. Scarani, C. Kurtsiefer, Phys. Rev. Lett. \textbf{103}, 153601 (2009)
\bibitem{PRAHetet2013}
G. H\'etet, L. Slodi\ifmmode \check{c}\else \v{c}\fi{}ka, N. R\"ock, R. Blatt, Phys. Rev. A \textbf{88}, 041804 (2013)
\bibitem{PRLZumofen2008}
G. Zumofen, N. M. Mojarad, V. Sandoghdar, M. Agio, Phys. Rev. Lett. \textbf{101}, 180404 (2008)
\bibitem{JourModOptLeuchs2012}
G. Leuchs, M. Sondermann, Journal of Modern Optics \textbf{60}, 36-42 (2012)
\bibitem{PhysRevAKochan1994}
P. Kochan, H. J. Carmichael, Phys. Rev. A \textbf{50}, 1700-1709 (1994)
\bibitem{NatPhysWrigge2008}
G. Wrigge, I. Gerhardt, J. Hwang, G. Zumofen, V. Sandoghdar, Nature Physics \textbf{4}, 60-66 (2008)
\bibitem{NJPTey2009}
M. K. Tey, G. Maslennikov, T. C. H. Liew, S. A. Aljunid, F. Huber, B. Chng, Z. Chen, V. Scarani, C. Kurtsiefer, New Journal of Physics \textbf{11}, 043011 (2009)
\bibitem{PRLSlodicka2010}
L. Slodi\ifmmode \check{c}\else \v{c}\fi{}ka, G. H\'etet, S. Gerber, M. Hennrich, R. Blatt, Phys. Rev. Lett. \textbf{105}, 153604 (2010)
\bibitem{PRAvanEnk2004}
S. J. van Enk, Phys. Rev. A \textbf{69}, 043813 (2004)
\bibitem{EurPhysJGolla2012}
A. Golla, B. Chalopin, M. Bader, I. Harder, K. Mantel, R. Maiwald, N. Lindlein, M. Sondermann, G. Leuchs, Eur. Phys. J. D \textbf{66}, 190-198 (2012)
\bibitem{OptCommTeo2011}
C. Teo, V. Scarani, Optics Communications \textbf{284}, 4485-4490 (2011)
\bibitem{PhysRevAMaiwald2012}
R. Maiwald, A. Golla, M. Fischer, M. Bader, S. Heugel, B. Chalopin, M. Sondermann, G. Leuchs, Phys. Rev. A \textbf{86}, 043431 (2012)
\bibitem{NatPhysMaiwald2009}
R. Maiwald, D. Leibfried, J. Britton, J. C. Bergquist, G. Leuchs, D. J. Wineland, Nature Physics \textbf{5}, 551-554 (2009)
\bibitem{LaserPhysLindlein2007}
N. Lindlein, R. Maiwald, H. Konermann, M. Sondermann, U. Peschel, G. Leuchs, Laser Physics \textbf{17}, 927-934 (2007)
\bibitem{ApplPhysBSondermann2007}
M. Sondermann, R. Maiwald, H. Konermann, N. Lindlein, U. Peschel, G. Leuchs, Applied Physics B \textbf{89}, 489-492 (2007)
\bibitem{JApplPhysBerkeland1998}
D. J. Berkeland, J. D. Miller, J. C. Bergquist, W. M. Itano, D. J. Wineland, J. Appl. Phys. \textbf{83}, 5025-5033 (1998)
\bibitem{PRAOlmschenk2007}
S. Olmschenk, K. C. Younge, D. L. Moehring, D. N. Matsukevich, P. Maunz, C. Monroe, Phys. Rev. A \textbf{76}, 052314 (2007)
\bibitem{PRABell1991}
A. S. Bell, P. Gill, H. A. Klein, A. P. Levick, Chr. Tamm, D. Schnier, Phys. Rev. A \textbf{44}, R20-R23 (1991)
\bibitem{ApplOptLeuchs2008}
G. Leuchs, K. Mantel, A. Berger, H. Konermann, M. Sondermann, U. Peschel, N. Lindlein, J. Schwider, Applied Optics \textbf{47}, 5570-5584 (2008)
\bibitem{NatureHwang2009}
J. Hwang, M. Pototschnig, R. Lettow, G. Zumofen, A. Renn, S. G\"otzinger, V. Sandoghdar, Nature \textbf{460}, 76-80 (2009)
%
\end{thebibliography}
\end{document}